\documentclass{cs23proc}

\usepackage{bm, amsmath}
\usepackage[
    colorlinks=true,
    linkcolor=blue,
    urlcolor=blue,
    citecolor=blue
]{hyperref}

\usepackage{placeins}
\allowdisplaybreaks

\makeatletter

\editors{Takeru Suzuki and the Cool Stars 23 Organizing Team}
\publisher{Zenodo}
\conference{The 23th Cambridge Workshop on Cool Stars, Stellar Systems, and the Sun (Cool Stars 23)}
\conferencedate{2026}

\defcitealias{Clark1973}{C73}
\defcitealias{Spiegel1992}{SZ92}
\defcitealias{Gough1998}{GM98}
\defcitealias{Matilsky2022}{M22}
\defcitealias{Matilsky2024}{M24}
\defcitealias{Matilsky2025a}{M25}
\defcitealias{Matilsky2026a}{M26}
\defcitealias{Korre2024}{KF24}
\defcitealias{ForgcsDajka2001}{FDP01}

\title{The relationship between solar and stellar tachoclines, dynamos, and spin-down}
\author{Loren I. Matilsky $^{1,2}$,
        Lydia Korre $^{3}$,
        Nicholas H. Brummell $^{1}$}

\affiliation{$^{1}$ {Department of Applied Mathematics, Baskin School of Engineering, University of California, Santa Cruz, CA 96064-1077, USA} \\
			 $^{2}$ U.S. National Science Foundation Astronomy and Astrophysics Postdoctoral Fellow \\
			 $^{3}$ Department of Applied Mathematics, University of Colorado, Boulder, CO 80309-0526, USA}

\shorttitle{Spin-down and solar/stellar tachoclines}
\shortauthors{Matilsky, Korre \& Brummell}

\providecommand{\sn}[2]{#1\times10^{#2}}

\providecommand{\cz}{_{\rm{cz}}}
\providecommand{\rz}{_{\rm{rz}}}

\providecommand{\dimm}{_{\rm{dim}}}

\providecommand{\pderiv}[2]{\dfrac{\partial#1}{\partial#2}}
\providecommand{\matderiv}[1]{\frac{D#1}{Dt}}
\providecommand{\pderivline}[2]{\partial#1/\partial#2}

\providecommand{\av}[1]{\left\langle#1\right\rangle}
\providecommand{\abs}[1]{\left\lvert#1\right\rvert}

\providecommand{\avt}[1]{\left\langle#1\right\rangle_{t}}

\providecommand{\five}{\ \ \ \ \ }

\providecommand{\andd}{\text{and}\five }
\providecommand{\where}{\text{where}\five }

\providecommand{\curl}{\nabla\times}
\providecommand{\Div}{\nabla\cdot}

\providecommand{\dotgrad}{\cdot\nabla}
\providecommand{\ugrad}{\bm{u}\dotgrad}

\providecommand{\e}{\hat{\bm{e}}}
\providecommand{\erad}{\e_r}
\providecommand{\etheta}{\e_\theta}
\providecommand{\ephi}{\e_\phi}

\providecommand{\ez}{\e_z}

\providecommand{\Omzero}{\Omega_0}
\providecommand{\twoOmzero}{2\Omega_0}
\providecommand{\Omzerovec}{\bm{\Omega}_0}

\providecommand{\ofr}{(r)}
\providecommand{\rprime}{{r^{\prime}}}
\providecommand{\ofrprime}{(\rprime)}

\providecommand{\cp}{c_{\rm{p}}}

\providecommand{\Dentr}{\Delta s}

\providecommand{\rhocz}{\rho\cz}

\providecommand{\tmpcz}{T\cz}

\providecommand{\bruntrz}{N\rz}

\providecommand{\gravcz}{g\cz}

\providecommand{\nucz}{\nu\cz}

\providecommand{\kappacz}{\kappa\cz}

\providecommand{\etacz}{\eta\cz}

\providecommand{\fluxnradcz}{F_{{\rm{nr,cz}}}}

\providecommand{\rhotilde}{\tilde{\rho}}
\providecommand{\tmptilde}{\tilde{T}}

\providecommand{\heatradtilde}{\tilde{Q}_{\rm r}}

\providecommand{\brunttilde}{\widetilde{N}}
\providecommand{\brunttildesq}{\widetilde{N}^2}

\providecommand{\gravtilde}{\tilde{g}}
\providecommand{\nutilde}{\tilde{\nu}}
\providecommand{\kappatilde}{\tilde{\kappa}}
\providecommand{\etatilde}{\tilde{\eta}}

\providecommand{\fluxscalarnradtilde}{\tilde{F}_{\rm{nr}}}

\providecommand{\prshat}{\hat{p}}
\providecommand{\entrhat}{\hat{s}}

\providecommand{\vecu}{\bm{u}}

\providecommand{\vecb}{\bm{B}}
\providecommand{\vecom}{\bm{\omega}}

\providecommand{\bpol}{\vecb_{\rm{pol}}}
\providecommand{\umer}{\vecu_{\rm{m}}}

\providecommand{\urad}{{u_r}}
\providecommand{\utheta}{{u_\theta}}
\providecommand{\uphi}{{u_\phi}}
\renewcommand{\uphi}{{u_\phi}}

\providecommand{\rhoumer}{\av{\rhotilde\umer}}

\providecommand{\brad}{B_r}
\providecommand{\btheta}{B_\theta}
\providecommand{\bphi}{B_\phi}

\providecommand{\umerprime}{\vecu_{\rm{m}}^\prime}

\providecommand{\amom}{\mathcal{L}}

\providecommand{\taurs}{\tau_{\rm{rs}}}
\providecommand{\taumc}{\tau_{\rm{mc}}}
\providecommand{\tauv}{\tau_{\rm{v}}}

\providecommand{\taums}{\tau_{\rm{ms}}}
\providecommand{\taumm}{\tau_{\rm{mm}}}

\providecommand{\ra}{{\rm{Ra}}}

\providecommand{\raf}{\ra_{\rm{f}}}

\providecommand{\pr}{{\rm{Pr}}}
\providecommand{\prm}{{\rm{Pr_m}}}
\providecommand{\ek}{{\rm{Ek}}}

\providecommand{\roc}{{\rm{Ro_c}}}
\providecommand{\rocsq}{{\rm{Ro_c^2}}}
\providecommand{\bu}{{\rm{Bu}}}

\providecommand{\di}{{\rm{Di}}}

\providecommand{\rin}{{r_{\rm in}}}

\providecommand{\rindim}{r_{\rm in}^*}
\providecommand{\routdim}{r_{\rm out}^*}

\providecommand{\rbcz}{r_{\rm bcz}}

\providecommand{\rc}{r_{\rm c}}

\providecommand{\rtach}{r_{\rm t}}

\providecommand{\rcdim}{{r_{\rm c}^*}}

\providecommand{\rsun}{R_\odot}

\providecommand{\rayleigh}{\texttt{Rayleigh}}

\renewcommand{\dimm}{^*}
\newcommand{\dimsq}{^{*2}}

\newcommand\lmout{\bgroup\markoverwith{\textcolor{red}{\rule[0.5ex]{2pt}{1pt}}}\ULon}
\newcommand\lkout{\bgroup\markoverwith{\textcolor{blue}{\rule[0.5ex]{2pt}{1pt}}}\ULon}
\newcommand\nbout{\bgroup\markoverwith{\textcolor{orange}{\rule[0.5ex]{2pt}{1pt}}}\ULon}

\renewcommand{\heatradtilde}{\tilde{Q}_{\rm rad}}

\abs{The solar tachocline, a thin shear layer separating the differentially rotating convective zone (CZ) from the underlying rigidly rotating radiative zone (RZ), remains a dynamical mystery. Under the influence of ``radiative spreading'' (the process by which meridional circulation, baroclinicity, and differential rotation burrow through a stably stratified fluid), the tachocline should have spread significantly by the current age of the Sun. Some unknown torque must therefore rigidify the whole solar interior below the CZ and keep the tachocline confined to a thin layer. At the same time, the spin-down process (through which cool stars shed angular momentum via surface magnetic torques) must extract angular momentum from the solar RZ, presenting a second mystery: solar spin-down must be communicated from the near-surface layers to the deep interior, all the while leaving the tachocline intact. In this work, we explicitly analyze the dynamics of radiative spreading in two 3D, spherical-shell, fully nonlinear fluid simulations of a solar-like CZ--RZ system, one without a magnetic field (hydrodynamic---HD) and one with a small random seed magnetic field (magnetohydrodynamic---MHD). We find that radiative spreading is unmitigated in the HD case (as expected), but in the MHD case, a self-excited dynamo not only confines the tachocline, but \textit{also} extracts angular momentum from the RZ, thereby communicating the spin-down downward. We thus speculate that solar and stellar tachoclines may be intimately linked to both the spin-down process and global dynamo.}

\begin{document}

\maketitle

\section{Introduction}
The solar tachocline is a thin shear layer inside the Sun at the base of the solar convective zone (CZ), where the strong CZ latitudinal differential rotation transitions rapidly to nearly rigid rotation in the underlying radiative zone (RZ; see \citealt{Brown1989,Howe2009,Basu2016}). The center of the tachocline lies at about $\rtach\approx0.7\rsun$ (where $\rsun=\sn{6.96}{10}$ cm is the radius of the Sun), which is also roughly the base of the CZ. The observed thickness of the tachocline, $\Delta$, is rather small, potentially as low as $\Delta\approx0.02\rsun$ \citep{Elliott1999}. Probably $\Delta$ is too small to be resolved, in which case the published values are really upper bounds, commensurate with the kernel width of the helioseismic inversion \citep{Howe2009}.

In any case, the observed thinness of the tachocline poses some major dynamical problems, many of which are still incompletely understood. The main issue is how the tachocline can remain thin, despite the action of ``radiative spreading'' (\citealt{Spiegel1992}; hereafter \citetalias{Spiegel1992}). Radiative spreading occurs because the CZ's baroclinicity (i.e., the horizontal temperature gradients which must balance variations of the centrifugal acceleration along the rotation axis in a type of ``thermal wind balance''; see \citealt{Matilsky2023}) spreads downward via thermal diffusion. This burrowing baroclinicity drives a large-scale meridional circulation (a bulk overturning motion in radius and latitude), and both the baroclinicity and the circulation spread downward significantly on long dynamical time-scales, despite being opposed by the RZ's strong stable stratification, which inhibits vertical motion (e.g., \citealt{Clark1973, Haynes1991}). Rotational shear (which is present everywhere in the solar CZ because of the strong latitudinal differential rotation) leads to the burrowing circulation transporting angular momentum and thus spreading the CZ's latitudinal shear inward, thickening the tachocline. Indeed, \citetalias{Spiegel1992} argued that, were the tachocline infinitesimally thin at solar birth, radiative spreading should have increased $\Delta=0$ to $\Delta\approx0.4\rsun$ after the solar age. 

If radiative spreading were unopposed, it should have eliminated the rigid rotation of the upper RZ, which is directly at odds with helioseismic observations. The observed thin tachocline thus implies the presence of a large-scale torque in the radiative interior that both rigidifies the deep interior and stops the tachocline from radiatively spreading (technically, one torque could confine the tachocline and another could rigidify the deep interior, but it is often assumed, rather intuitively, that both torques arise from the same mechanism). The theoretical origins of this torque are typically thought to be hydrodynamic (HD; e.g., \citealt{Spiegel1992,Kumar1997,Kim2007,Garaud2025}) or magnetohydrodynamic (MHD; e.g., \citealt{Gough1998,ForgcsDajka2001,ForgcsDajka2002,Garaud2002,ForgcsDajka2004,Strugarek2011a,Wood2012,AcevedoArreguin2013,Barnabe2017,Wood2018,Matilsky2024}). Here, we focus on the torque caused by the Maxwell stresses from large-scale nonaxisymmetric dynamo modes, which are first generated in the CZ and then diffuse into the upper RZ via a type of skin effect (e.g., \citealt{Matilsky2024}). These inward-diffusing Maxwell stresses can stop the radiative spreading and confine the tachocline in a process we refer to as the ``dynamo confinement scenario.''

The outer CZ-envelope dynamos in cool stars also cause them to spin down. The magnetic field lines from the dynamo are drawn outward by the ionized stellar wind and these field lines act like a long moment arm that applies a significant decelerating torque to the star at its surface (e.g., \citealt{Parker1955}). In theory, these torques do not directly affect the deeper interior (i.e., the RZ) and we thus might expect expect that cool-star RZs rotate very fast, inheriting the very rapid rotation from stellar birth. In the Sun, however, this is not what helioseismology observes: instead, the upper solar RZ rotates at a rate commensurate with the bulk rotation rate of the CZ envelope. The relatively slow rotation of the upper solar RZ and simultaneous presence of a thin tachocline present an interesting paradox. On the one hand, the tachocline should act like a sort of ``dynamical barrier'' between CZ and RZ, with little angular momentum transport occurring across it. That is because clearly the CZ's differential rotation, with its associated angular momentum anomaly, has not been able to spread downward. On the other hand, the fact that the RZ rotates slowly at the bulk rate of the rest of the star suggests that the spin-down torques \textit{do} communicate downward, across the tachocline barrier.
 
 

In this work, we examine the tachocline/spin-down paradox by quantifying the finer details of radiative spreading in two 3D, spherical-shell, fully nonlinear, anelastic fluid simulations using the {\rayleigh} code. One simulation is purely HD with no magnetic field and the other is MHD, initiating a self-excited dynamo from a small random seed field. Below, we assess how well the fluid equations obey the dynamics of radiative spreading as outlined by \citetalias{Spiegel1992} and discuss the role these dynamics play in tachocline formation and spin-down. In particular, we show the torque balance for circulation burrowing (including the influence of magnetism in the MHD case) and the thermal wind balance (e.g., \citealt{Matilsky2023}) during representative time intervals as the radiative spreading process unfolds. 

\section{Methods}
This proceedings paper is a complement to a much longer article in the astrophysical journal (\citealt{Matilsky2026a}; hereafter \citetalias{Matilsky2026a}). For more details on the numerical setup, see \citetalias{Matilsky2026a}. This proceedings is also part of a longer series of papers \citep{Matilsky2021,Matilsky2022,Matilsky2024,Matilsky2025a,Matilsky2026a} exploring the dynamo confinement scenario. Here, we focus specifically on how the dynamo confinement scenario interacts with radiative spreading, with the goal of understanding the communication of spin-down across solar and stellar tachoclines. 

\begin{figure*}[ht!]
	\centering
	\includegraphics[width=\textwidth]{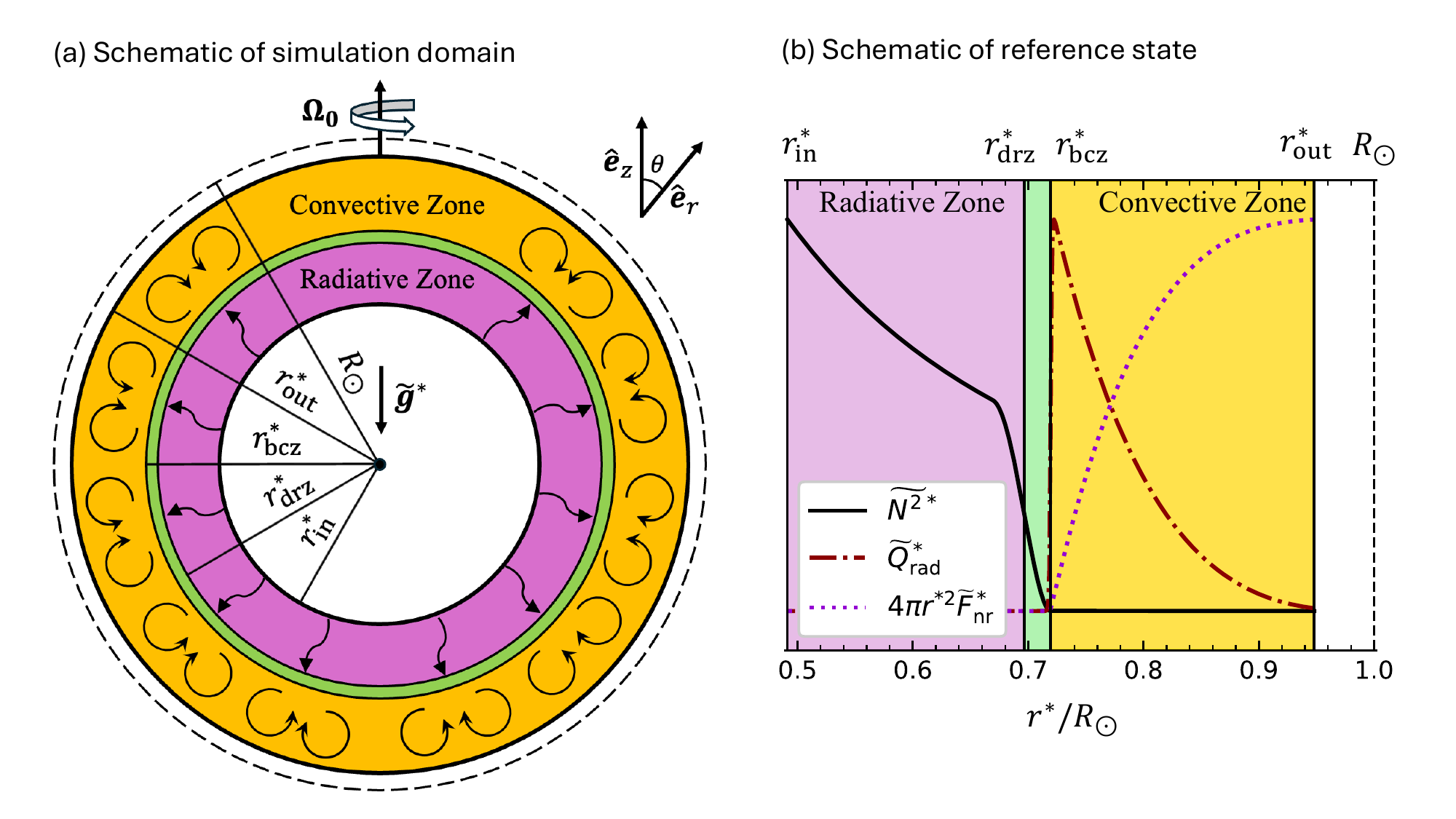}
	\caption{(a) Schematic of our three-dimensional spherical-shell simulation domain. The yellow region depicts the CZ with bulk overturning fluid motions represented by circular arrows. The CZ lies atop a thin stably stratified region (green) with significant nonlinear effects from overshoot and a deep RZ (purple) where the motion is essentially linear, represented by curvy ``radiation" arrows. (b) Schematic of the reference state which enforces the geometry of panel (a). Convection is forced in roughly the bottom third of the CZ by a prescribed internal heating $\heatradtilde\dimm$, which causes an accompanying ``nonradiative" energy flux $4\pi r\dimsq\fluxscalarnradtilde\dimm$ that convection and conduction must carry in a steady state to maintain thermal equilibrium (see Equation \ref{eq:fnrad}). The stable stratification is enforced through the positive squared background buoyancy frequency $\brunttilde\dimsq$ in the RZ. These are the shapes of the profiles used for both the HD and MHD cases, and have been normalized to all lie on the same scale. The dashed circle (panel a) and dashed line (panel b) mark the location of the true solar surface. This figure and caption were adapted from Figure 1 of \citetalias{Matilsky2026a}. }
	\label{fig:schematic}
\end{figure*}

We consider two simulations from \citetalias{Matilsky2026a}---there referred to as ``Cases H6 and M6''---which we here refer to the ``HD case'' and ``MHD case,'' respectively. The simulation domain is a 3D spherical shell, pictured in Figure \ref{fig:schematic}, with dimensional inner radius $\rindim$ and outer radius $\routdim$. At the midpoint radius $\rcdim$, the shell is separated into a CZ above (with convection driven by a CZ-only background internal heating function $\heatradtilde^*$ meant to mimic radiative heating) and a stably stratified RZ below (with convective stability enforced through a positive squared background buoyancy frequency $\brunttilde\dimsq$). Asterisks denote explicitly dimensional quantities whereas the lack of asterisks denote nondimensional quantities. We use spherical coordinates $r\dimm$ (dimensional radius), $\theta$ (colatitude), and $\phi$ (azimuth angle), as well as cylindrical coordinates $\lambda\dimm= r\dimm\sin\theta$ (dimensional cylindrical radius, or moment arm) and $z\dimm= r\dimm\cos\theta$ (dimensional axial coordinate). We denote unit vectors by $\e$, with a subscript indicating the local coordinate direction of the unit vector (for example, $\ephi$ is the azimuthal unit vector). The evolution equations are solved in a rotating frame with constant background angular velocity $\Omzerovec=\Omzero\ez$. 

We use the {\rayleigh} code \citep{Featherstone2016a,Matsui2016,Featherstone2021} to solve the anelastic (M)HD nondimensional equations of motion with respect to time $t$ for the vector velocity $\vecu$, the vector magnetic field $\vecb$, the pressure perturbation (away from the spherically symmetric background state) $\hat{p}$, and the entropy perturbation $\hat{s}$: 
\begin{align}
	\Div(\rhotilde\vecu) &\equiv 0\label{eq:cont},\\
	\Div \vecb &\equiv 0\label{eq:divb0},
\end{align}
\begin{subequations}
\begin{align}\label{eq:mom}
    \rhotilde\left(\matderiv{\vecu}\right) = &-\rhotilde\ez\times\vecu-\rhotilde\nabla\left(\frac{\prshat}{\rhotilde} \right) +\rocsq \rhotilde\, \gravtilde \entrhat\erad \nonumber\\
    & +\sqrt{\frac{\pr}{\raf}}\roc \Div (D_{ij}\e_i\e_j)\nonumber\\
    & +(\curl\vecb)\times\vecb,\\
    \where D_{ij}\dimm =\ &\left\{ 2\rhotilde\,\nutilde \left[e_{ij} - \frac{1}{3}(\nabla\cdot\vecu) \delta_{ij} \right]\right\}\label{eq:vstress}\\
    \andd e_{ij} =\ &\frac{1}{2}\left(\pderiv{u_i}{x_j} + \pderiv{u_j}{x_i} \right),\label{eq:ratestrain}
\end{align}
\end{subequations}
\begin{align}\label{eq:heat}
	\rhotilde\tmptilde \matderiv{\entrhat} = &- \frac{\sigma^2}{\pr\rocsq} \rhotilde\tmptilde \frac{\brunttildesq}{\gravtilde} u_r+ \frac{\roc}{\sqrt{\pr\raf}} \heatradtilde\nonumber\\
    &+\frac{\roc}{\sqrt{\pr\raf}}\Div(\rhotilde \tmptilde \kappatilde \nabla \entrhat)  \nonumber\\
	&+ \sqrt{\frac{\pr}{\raf}}\frac{\di}{\roc}\left( D_{ij}e_{ij} + \frac{\etatilde}{\prm} |\curl\vecb|^2\right),
\end{align}
and
\begin{align}\label{eq:ind}
	\pderiv{\vecb}{t} = \curl(\vecu\times\vecb) - \sqrt{\frac{\pr}{\raf}}\frac{\roc}{\prm} \curl(\etatilde\curl\vecb).
\end{align}	

Here, $D/Dt=\pderivline{}{t} + \ugrad$ is the material derivative, $\delta_{ij}$ is the Kronecker delta, and repeated indices are summed over the three coordinate directions. The tildes denote a fixed, spherically symmetric ``reference'' or ``background'' state, consisting of the nondimensional density $\rhotilde\ofr$, temperature $\tmptilde\ofr$, gravity $\gravtilde\ofr$, kinematic viscosity $\nutilde\ofr$, thermometric conductivity $\kappatilde\ofr$, magnetic diffusivity $\etatilde\ofr$, squared buoyancy frequency $\brunttildesq\ofr$, and internal CZ-only heating $\heatradtilde\ofr$.

In the equations of motion, the coordinates and fields have been nondimensionalized using the following system of units:
\begin{subequations}\label{eq:nond}
\begin{align}
[\nabla\dimm] &= H^{-1},\\
[\partial/\partial t\dimm] &= \twoOmzero,\\
[\vecu\dimm] &= \twoOmzero H,\\
[\vecb\dimm] &= \sqrt{4\pi\rhocz}(\twoOmzero H),\\
[\prshat\dimm] &= \rhocz(\twoOmzero H)^2,\\
\andd [\entrhat\dimm] &=\Dentr = \frac{\fluxnradcz H}{\rhocz\tmpcz\kappacz}.\label{eq:deltas}
\end{align}
\end{subequations}
Here, the ``cz'' and ``rz'' subscripts on a background-state quantity denote the volume-average of a dimensional background profile over the CZ or RZ, respectively. The CZ or RZ dimensional depth is $H=\routdim-\rcdim=\rcdim-\rindim$. The dimensional ``nonradiative energy flux,''
\begin{equation}\label{eq:fnrad}
\fluxscalarnradtilde\dimm(r\dimm) =\frac{1}{r\dimsq}\int_{\rbcz\dimm}^{r\dimm}\heatradtilde\dimm\ofrprime, \rprime^2d\rprime
\end{equation}
represents the energy flux that convection and conduction must carry in the final statistically steady state (see \citealt{Featherstone2016a} and Figure \ref{fig:schematic}b). For more details on this largely solar-like reference state and its nondimensionalization, see \citetalias{Matilsky2026a}. 

The control parameters appearing in Equations \eqref{eq:cont}--\eqref{eq:ind} are:
\begin{subequations}\label{eq:control}
\begin{align}
     \pr &= \frac{\nucz}{\kappacz} \five \text{(thermal Prandtl number)},\label{eq:controlpr}\\
     \prm &= \frac{\nucz}{\etacz} \five \text{(magnetic Prandtl number)},\\
     \raf &= \frac{\gravcz H^3}{\nucz\kappacz}\left(\frac{\Dentr}{\cp}\right)= \frac{\fluxnradcz \gravcz H^4}{\rhocz\tmpcz\cp\nucz\kappacz^2}\nonumber\\
     &\five \text{(flux-based Rayleigh number)},\\
      \roc &= \frac{\sqrt{\gravcz(\Dentr/\cp)}}{\twoOmzero} =\ek\sqrt{\frac{\ra}{\pr}}\nonumber\\
      &\five \text{(convective Rossby number)}\label{eq:controlroc},\\
      &\nonumber\\
      &\text{and}\nonumber\\
       \sigma &= \left(\frac{\bruntrz}{\twoOmzero}\right)\sqrt{\frac{\nucz}{\kappacz}}=\sqrt{\bu\pr}\nonumber\\
       &\five\text{(sigma  parameter).}\label{eq:controlsigma}
\end{align}
\end{subequations}

Here, $\cp$ is the specific heat at constant pressure. In Equations \eqref{eq:controlroc} and \eqref{eq:controlsigma} we have implicitly defined the Ekman number,
\begin{align}\label{eq:controlek}
    \ek&= \frac{\nucz}{\twoOmzero H^2}
\end{align}
and the buoyancy number,
\begin{align}\label{eq:controlbu}
    \bu=\frac{\bruntrz^2}{4\Omzero^2}.
\end{align}

The five independent parameters in Equations \eqref{eq:control} fully characterize the system and we thus treat $\ek$ and $\bu$ as dependent parameters. The ``dissipation number,''
\begin{equation}\label{def:di}
\di=\frac{\gravcz H}{\cp\tmpcz},
\end{equation} 
is a property of the reference state (see Appendix A of \citetalias{Matilsky2026a}). 

For the two cases considered here: $\pr=0.25$, $\prm=4$, $\raf=\sn{5.62}{5}$, $\roc=0.4$, $\sigma=2.88$, $\ek=\sn{2.67}{-4}$, $\bu=33.1$, and $\di=1.72$.  

In each case, convection is initialized with small random seed perturbations in $\hat{s}$. In the MHD case, the dynamo is initialized from a small random seed field in $\vecb$. All other quantities are initialized to zero. At each boundary, impenetrable, stress free, potential-field-matching, and fixed-entropy-gradient conditions are used (see \citetalias{Matilsky2026a}). 

\section{Angular momentum transport}\label{sec:amom}
\begin{figure}
	\centering
	\includegraphics[width=\linewidth]{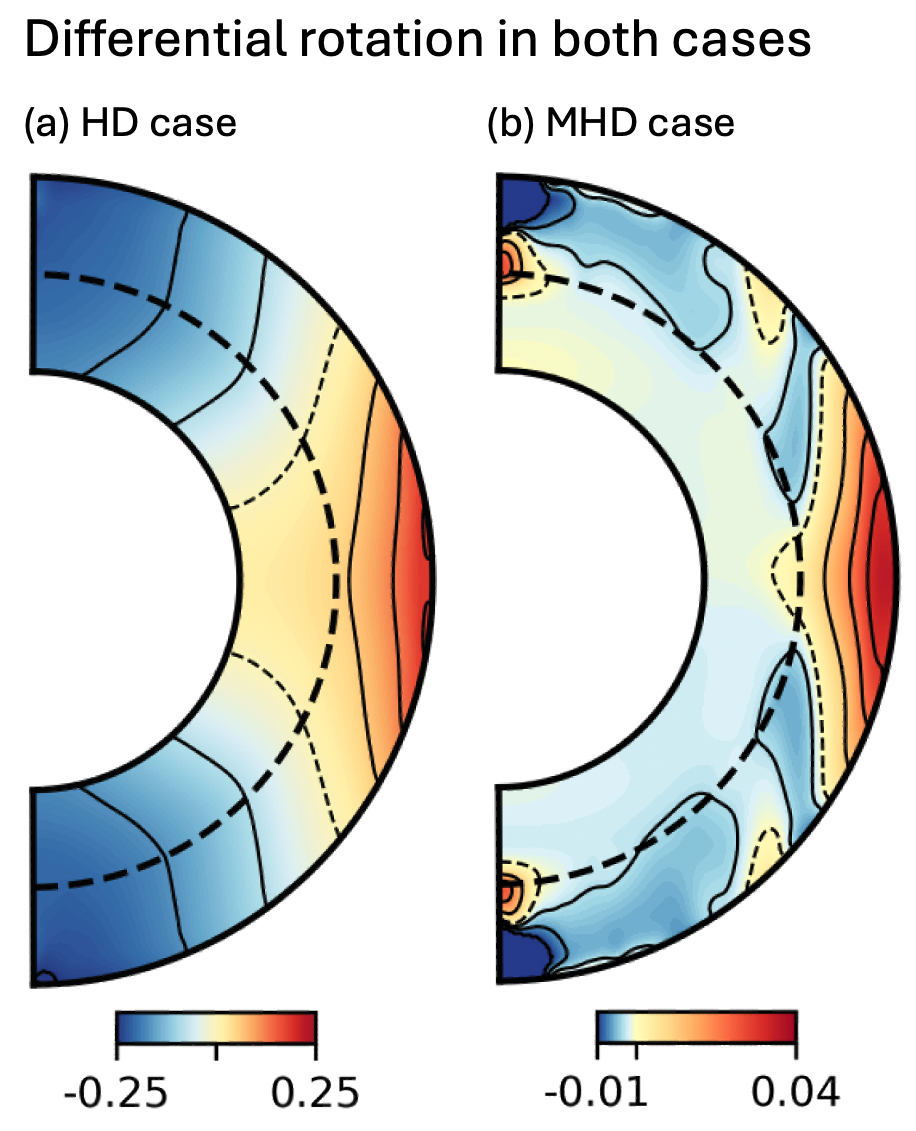}
	\caption{\textbf{A dynamo confines a thin solar-like tachocline and causes RZ spin-down.} We show the ``differential rotation'' $\avt{2\Omega-1}$ (see Equation \ref{eq:diffrot}) for (a) the HD case and (b) the MHD case, temporally averaged over each simulation's equilibrated state (the angular brackets with a ``$t$'' subscript denote a temporal average). Negative values (blue tones) are normalized separately from positive values (red tones), each with equally spaced contours marked by black curves (the zero contour is dashed). The thick dashed semicircles mark the CZ--RZ boundary $r=\rc$. This figure was adapted from Figure 4 of \citetalias{Matilsky2026a}. }
	\label{fig:DR}
\end{figure}

We define the nondimensional time-dependent azimuthally averaged rotation rate $\Omega$ in the meridional plane by
\begin{equation}\label{eq:omega}
\Omega(r,\theta,t) = \frac{1}{2} + \frac{\av{\uphi}}{\lambda},
\end{equation}
where the angular brackets denote an instantaneous azimuthal average. Note that for our choice of units given in Equation \ref{eq:nond}, the frame rotation rate is simply $1/2$. In Figure \ref{fig:DR}, we show the ``differential rotation'' for both the HD and MHD cases, defined by
\begin{align}\label{eq:diffrot}
2\Omega-1 = \frac{2\av{\uphi}}{\lambda} = \frac{\Omega\dimm}{\Omzero}-1.
\end{align}
In the HD case, the differential rotation has spread all the way into the RZ, but in the MHD case, the RZ rotates rigidly and there is a confined tachocline. Furthermore, the RZ has been slightly spun down to rotate less rapidly than the frame rate (i.e., $\Omega<1/2$ in most of the RZ). 

To understand the MHD case's simultaneous confinement of a tachocline and spin-down of the deep RZ, we examine the time-dependent angular momentum transport (torque) equation in both cases. We define the specific angular momentum density of the fluid as: 
\begin{equation}\label{eq:amom}
	\amom= \lambda\left(\frac{\lambda}{2} + \av{\uphi}\right)=\lambda^2\Omega.
\end{equation} 
Multiplying the $\phi$-component of the azimuthal mean of Equation \eqref{eq:mom} by $\lambda$ yields the evolution equation for $\amom$:
\begin{subequations}\label{eq:torque}
	\begin{align}
		\rhotilde\pderiv{\amom}{t}=&
		-\Div[\rhotilde (\lambda \av{u^\prime_\phi \umerprime} 
		+\amom\av{\umer} -\ek\nutilde\lambda^2\nabla\Omega)\nonumber\\
		&- \lambda (\av{ B_\phi^\prime\bpol^\prime} +  \av{\bphi}\av{\bpol} )]\\
		=\ & \taurs + \taumc + \tauv + \taums + \taumm,
	\end{align}
\end{subequations}
where we have defined the appropriate time-dependent torque densities:
\begin{subequations}\label{eq:torques}
	\begin{align}
		\taurs = &\ -\Div (\rhotilde (\lambda \av{u^\prime_\phi \umerprime})\nonumber\\
		&\ \text{(Reynolds-stress torque)}, \label{eq:taurs}\\
		\taumc =  &\ -\rhoumer\cdot\nabla\amom \nonumber\\
		&\ \text{(meridional-circulation torque)},\label{eq:taumc}\\
		\tauv =  &\ \ek \Div(\rhotilde\nutilde\lambda^2\nabla\Omega)\nonumber\\ 
		&\ \text{(viscous torque)},\label{eq:tauv}\\
		\taums =  &\ \Div( \lambda \av{ B_\phi^\prime\bpol^\prime})\nonumber\\ 
		&\ \text{(Maxwell-stress torque)}, \label{eq:taums}\\
		\andd \taumm = &\ \Div( \lambda \av{\bphi}\av{\bpol})\nonumber\\
		&\ \text{(mean magnetic torque)}.\label{eq:taumm}
	\end{align}
\end{subequations}
Here, $\bpol= \brad\erad+\btheta\etheta$ is the vector poloidal (meridional) magnetic field, $\umer=\urad\erad+\utheta\etheta$ is the vector meridional velocity (i.e., $\av{\umer}$ is the meridional circulation), and the primes denote deviations from the instantaneous azimuthal average. 

\begin{figure*}
	\centering
	\includegraphics[width=\textwidth]{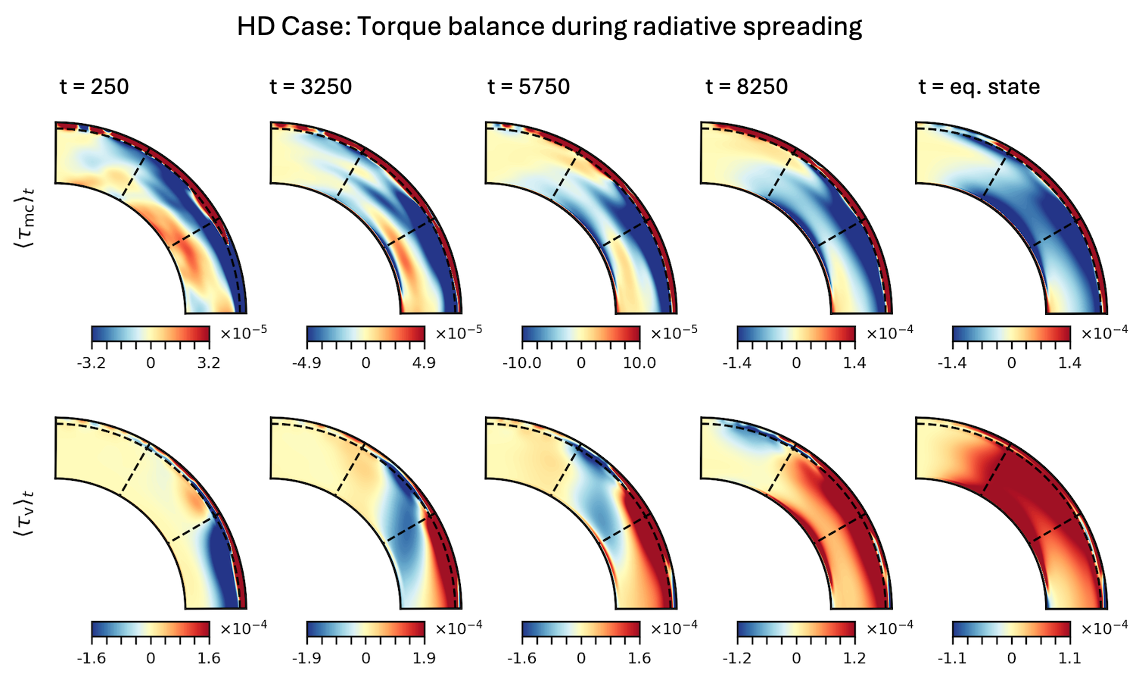}
	\caption{\textbf{Burrowing circulation causes RZ spin-down during radiative spreading}. We show the dominant torque densities from Equation \eqref{eq:torques}, which are due to the burrowing meridional circulation ($\avt{\taumc}$, top row) and the viscosity ($\avt{\tauv}$, bottom row) at representative times in the HD case, symmetrized about the equator, for the RZ only ($\rin\leq r\leq\rc$). Very early on ($t=250$), both torques are negative in large portions of the RZ, leading to some initial spin-down. Very quickly, however, the burrowing-circulation torque becomes negative in most of the RZ (thus causing most of the spin-down), whereas the viscous torque becomes positive. The internal dashed quarter-circles at $r=\rc-0.1$ mark the approximate depth of the overshoot layer (see the green region in Figure \ref{fig:schematic}). For the first four columns, each torque has been temporally averaged over an interval of length roughly $\Delta t\approx500$, centered at the indicated time. In the final column, the averaging interval is over the full equilibrated state, which for the HD case is $t=(32500,43600)$.  }
	\label{fig:Torque_HD}
\end{figure*}

Figure \ref{fig:Torque_HD} shows the dominant torque balances during radiative spreading for the HD case. The dynamics are complicated slightly by viscous effects, but in general, $\avt{\taumc}$ is negative in most of the RZ, thus causing most of the spin-down. Note, however, that essentially the only positive torque in the RZ (which must give rise to the HD case's fast equator in the deep RZ, which is evident from Figure \ref{fig:DR}a) is from the viscosity, i.e., $\avt{\tauv}>0$ at low latitudes for early times. This may be understood from geometry and the Taylor--Proudman theorem: radiative spreading tends to spread differential rotation primarily along contours parallel to the rotation axis (to enforce $\pderivline{\Omega}{z}\approx0$). The contour of zero differential rotation ($\Omega=1/2$; thin dashed contours in Figure \ref{fig:DR}) is biased toward low latitudes, which is a consequence of angular momentum conservation during latitudinal transport. Thus, the CZ's negative high-latitude differential rotation ($\Omega<1/2$) takes up a greater latitude range than the positive low-latitude differential rotation (fast equator). The cylindrical extension of the CZ's differential rotation is thus largely negative, giving rise to a negative $\avt{\taumc}$ and subsequent RZ spin-down. Because the HD case's differential rotation (in the CZ at least) is qualitatively rather similar to the solar differential rotation (e.g., \citealt{Howe2009}), the same behavior is expected in the Sun: i.e., we expect that radiative spreading from the Sun's burrowing circulation should cause spin-down.

\begin{figure*}
	\centering
	\includegraphics[width=\textwidth]{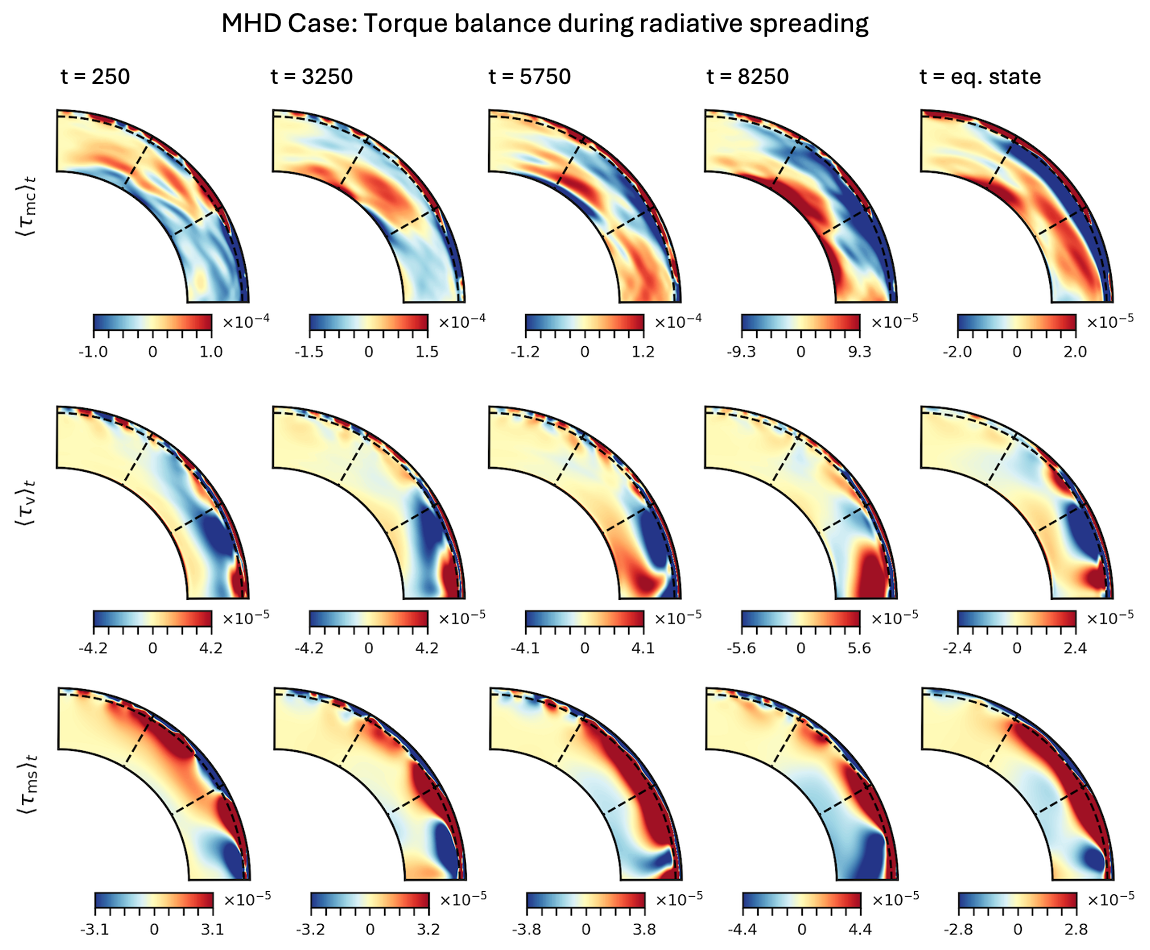}
	\caption{\textbf{The Maxwell stresses confine the tachocline and cause spin-down of the deep RZ}. Like Figure \ref{fig:Torque_HD}, but for the MHD case, now including the Maxwell-stress torque $\avt{\taums}$ (third row). The radiative spreading dynamics are now more complicated, but largely: the burrowing-circulation torque $\avt{\taumc}$ causes most of the spin-down initially (viscous effects are relatively weaker in the MHD case), but closer to the equilibrated state, the circulation only spins down the upper RZ and spins the lower RZ \textit{up}; the Maxwell-stress torque from the dynamo arranges itself to largely oppose the burrowing circulation, thus confining a tachocline and spinning \textit{down} the lower RZ. For the MHD case, the equilibrated state corresponds to $t=(10500, 46600)$. }
	\label{fig:Torque_MHD}
\end{figure*}

In the MHD case, the Maxwell stresses (which were shown to confine the tachocline; e.g., \citealt{Matilsky2022,Matilsky2024,Matilsky2026a}) also play a significant role in the torque balance (Figure \ref{fig:Torque_MHD}). The Maxwell-stress torque $\avt{\taums}$ primarily works to locally oppose the burrowing-circulation torque $\avt{\taumc}$ (and secondarily also works to oppose the viscous spreading of the fast equator, which is strongest at low latitudes). This results in a nearly complete cessation of both types of spreading (and thus confinement of a tachocline) in the equilibrated state, for which the spin-down from radiative spreading (negative $\avt{\taumc}$) is concentrated in a thin layer of the upper RZ just below the overshoot layer (i.e., in the tachocline). Maxwell stresses, however, also tend to rigidify any region connected by poloidal field lines (e.g., \citealt{Ferraro1937,Mestel1987}). Evidently this latter effect is so strong that the Maxwell stresses drag the whole deep RZ (below the tachocline) to a rigid rotation rate, which is negative owing to the radiative spreading. In the torque balance, this shows up as a \textit{negative} $\avt{\taums}$ in the deep RZ, which is then countered by a \textit{positive} torque $\avt{\taumc}$ from the (no-longer burrowing) circulation. 

In summary, the tachocline/spin-down paradox is largely resolved by the dynamo confinement scenario. If the dynamics of our MHD are representative of the Sun, then the picture we have is the following. Radiative spreading (or burrowing circulation) works to spin down the tachocline region by enforcing the Taylor--Proudman constraint ($\pderivline{\Omega}{z}=0$), which imprints the largely-negative high-latitude CZ differential rotation along cylinders into the RZ. The Maxwell stresses from the dynamo efficiently counter the radiative spreading (confining the tachocline), but in doing so, propagate the spin-down deep below the tachocline by rigidifying the rest of the RZ.  


\section{Thermal wind balance}
True radiative spreading, according to \citetalias{Spiegel1992}, consists of a particular set of dynamical balances holding in the HD equations of motion. The first of these ($\Div(\rhoumer)\equiv0$) is enforced by the anelastic approximation and the second---the spreading of differential rotation being dominated by the meridional-circulation torque---we verified for each case in the last section. In this section, we focus on the third balance (for the MHD case only), which is a type of ``thermal wind balance'' appropriate for stars (e.g., \citealt{Matilsky2023}). (The fourth balance---advective--diffusive balance in the heat equation---we do not consider here for brevity.) Taking the $\phi$-component of the curl of the specific momentum equation (Equation \ref{eq:mom} divided by $\rhotilde$) results in 
\begin{align}\label{eq:twb}
\rhotilde\pderiv{\av{\omega_\phi}}{t} = \rhotilde\lambda\pderiv{\Omega^2}{z} - \frac{\rocsq\rhotilde\gravtilde}{r}\pderiv{\entrhat}{\theta} + \dots,
\end{align}
where $\vecom=\curl\vecu$ is the vector vorticity and the ellipses denote other terms that are expected to be higher order. When the right-hand side of this equation is close to zero, the meridional circulation (which has a one-to-one correspondence with $\av{\omega_\phi}$) is only weakly evolving and the system is said to be in ``thermal wind balance.''

\begin{figure*}
	\centering
	\includegraphics[width=\textwidth]{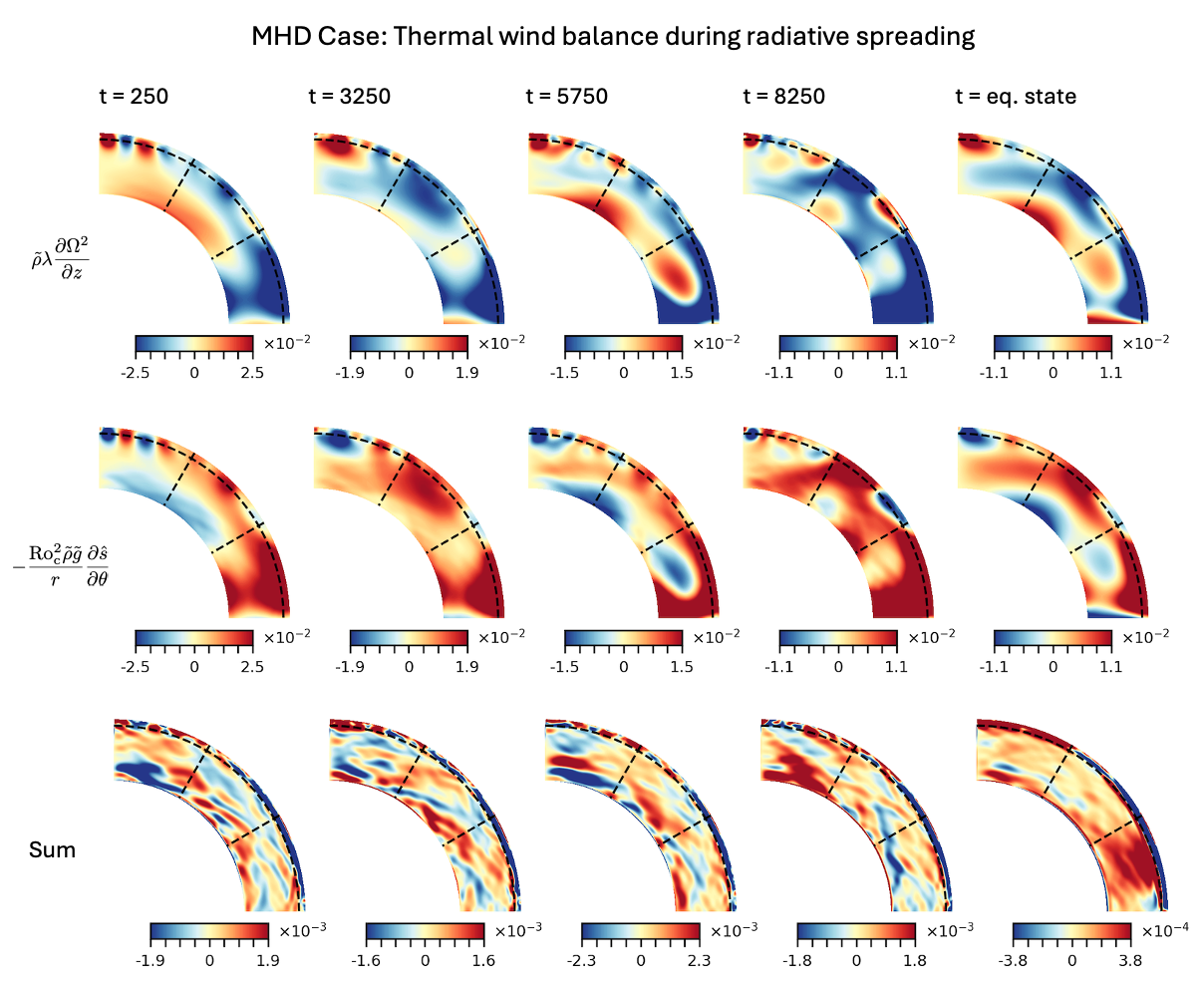}
	\caption{\textbf{Thermal wind balance holds even for a confined tachocline}. Like Figure \ref{fig:Torque_MHD} for the MHD case (with the same time intervals considered), but showing the two terms in the thermal wind balance Equation \eqref{eq:twb} and their sum. The structure of either term is rather complicated (and evolves significantly with time) but overall thermal wind balance holds very well at all times, with deviations from the balance being about 5--10\% while the system is evolving and about 3\% in the equilibrated state. Recall that for the MHD case, the equilibrated state corresponds to $t=(10500, 46600)$.}
	\label{fig:TWB_MHD}
\end{figure*}

Figure \ref{fig:TWB_MHD} shows the time-evolving thermal wind balance during radiative spreading for the MHD case, averaging in time over the same intervals as plotted for the torques in Figure \ref{fig:Torque_MHD}. Because Taylor--Proudman balance ($\pderivline{\Omega}{z}\equiv0$) induced by radiative spreading is a special form of thermal wind balance, and the Maxwell-stress torque was shown to stop the radiative spreading deep in the RZ, we might expect the MHD case to violate thermal wind balance, especially deeper in the RZ. Somewhat surprisingly, however, Figure \ref{fig:TWB_MHD} shows that this is emphatically not the case. Thermal wind balance holds at \textit{all} times, with violations on the order of only 10\% while the system is still evolving and 3\% in the equilibrated state. These results suggest that the dynamo confinement scenario can counter radiative spreading (by producing an additional confining torque term in the angular momentum equation) \textit{without} violating the other time-independent dynamical balances considered by \citetalias{Spiegel1992}. Again if these dynamics are consistent with the real solar interior, we expect a rigidly rotating RZ (weak differential rotation) to contain weak baroclinicity and meridional circulation as well. In other words, the tachocline is not only expected to be a transition in \textit{shear}, but a transition in latitudinal heat distribution and meridional circulation. (To verify this more precisely, we would need to assess advective--diffusive balance in the heat equation as well.)

\section{Discussion}
We have shown that the dynamo confinement scenario---the process through which the nonaxisymmetric modes from a convectively driven dynamo diffuse into the RZ and confine a tachocline by rigidifying the whole RZ---is also capable of extracting angular momentum from the RZ and thus potentially resolves the tachocline/spin-down paradox. RZ spin-down occurs through the combined effects of radiative spreading and Maxwell stresses. These results were already hinted at in the discussion of \citetalias{Matilsky2026a} (see Figure 14 from that paper), but here we have elucidated two more important points. First, the fundamental reason that radiative spreading is expected to cause spin-\textit{down} (as opposed to also imprinting the fast equator significantly) is due to the geometry of the solar differential rotation (for which the top of the RZ touches mostly high-latitude, slowly rotating portions of the CZ) and radiative spreading's tendency to enforce the Taylor--Proudman constraint. Second, despite necessarily violating the Taylor--Proudman constraint significantly to confine a tachocline, the Maxwell stresses do \textit{not} violate thermal wind balance, even while the system is evolving. If advective--diffusive balance in the heat equation is similarly not violated, we thus expect radiative spreading to unfold largely according to the original arguments of \citetalias{Spiegel1992}, even in the presence of a tachocline. In other words, whatever stops radiative spreading in the Sun and enforces rigid rotation of the RZ---be it a dynamo confinement scenario or some other mechanism---should, in principle, also stop the inward spread of baroclinicity and burrowing circulation. If it is the same physical process that confines the tachocline and rigidifies the whole RZ---as is the case for the dynamo confinement scenario---then we expect radiative spreading to extract angular momentum from the tachocline region (spinning it down) and the confining torque to extract angular momentum from deeper regions as well to rigidify the system. 

The natural next step in this research is to determine whether these newly discovered dynamics are to be expected in cool stars other than the Sun. To do this, one should repeat calculations like those presented here under varying levels of rotational constraint ($\roc$) and strength of RZ stable stratification ($\bu$) appropriate for other stellar interiors. If the same dynamics are found to occur, then we might expect the universal presence of tachoclines in cool stars (which is often assumed, though with little physical basis), and radiative cores that rotate at the relatively slow bulk rotation rate of the star, rather than the rapid rotation rate from stellar birth. If universal, the dynamics discussed here thus suggest an intimate synergy between tachoclines, dynamos, and spin-down in cool stars. 

\section*{Acknowledgments}
{We thank the organizers of Cool Stars 23 for a highly successful conference and timely opportunity to present this work. We thank P. Garaud, J. Toomre, N. Featherstone, B. Hindman, C. Blume, J. Pedlosky, V. Skoutnev, and A. Strugarek for helpful discussions. This work was primarily supported by the COFFIES DRIVE Science Center (NASA grant 80NSSC22M0162), with additional support from NSF award AST-2202253 and NASA grants 80NSSC18K1125, 80NSSC18K1127, 80NSSC19K0267, 80NSSC20K0193, 80NSSC21K0455, 80NSSC24K0270 and 80NSSC24K0125. Computational resources were provided by the NASA Advanced Supercomputing (NAS) Division at Ames Research Center. {\rayleigh} is supported by the Computational Infrastructure for Geodynamics (CIG) through NSF awards NSF-0949446 and NSF-1550901. For reproducibility, the {\rayleigh} input files and final checkpoints for both simulations considered here are publicly accessible at \url{https://doi.org/10.5281/zenodo.18275692}.}

\bibliographystyle{cs23proc}
\bibliography{example.bib}

@article{AcevedoArreguin2013,
	doi = {10.1093/mnras/stt1065},
	url = {https://doi.org/10.1093/mnras/stt1065},
	year = {2013},
	month = jul,
	publisher = {Oxford University Press ({OUP})},
	volume = {434},
	number = {1},
	pages = {720--741},
	author = {L. A. Acevedo-Arreguin and P. Garaud and T. S. Wood},
	title = {Dynamics of the solar tachocline {\textendash} {III}. {Numerical} solutions of the {Gough} and {McIntyre} model},
	journal = {MNRAS}
}

@article{Barnabe2017,
	doi = {10.1051/0004-6361/201630178},
	url = {https://doi.org/10.1051/0004-6361/201630178},
	year = {2017},
	month = apr,
	publisher = {{EDP} Sciences},
	volume = {601},
	pages = {A47},
	author = {Roxane Barnab{\'{e}} and Antoine Strugarek and Paul Charbonneau and Allan Sacha Brun and Jean-Paul Zahn},
	title = {Confinement of the solar tachocline by a cyclic dynamo   magnetic field},
	journal = {A\&A}
}

@article{Basu2016,
	doi = {10.1007/s41116-016-0003-4},
	url = {https://doi.org/10.1007/s41116-016-0003-4},
	year = {2016},
	month = aug,
	publisher = {Springer Science and Business Media {LLC}},
	volume = {13},
	number = {1},
	author = {Sarbani Basu},
	title = {Global seismology of the {Sun}},
	journal = {LRSP}
}

@article{Brown1989,
	doi = {10.1086/167727},
	url = {https://doi.org/10.1086/167727},
	year = {1989},
	month = aug,
	publisher = {American Astronomical Society},
	volume = {343},
	pages = {526},
	author = {Timothy M. Brown and Jorgen Christensen-Dalsgaard and Wojciech A. Dziembowski and Philip Goode and Douglas O. Gough and Cherilynn A. Morrow},
	title = {Inferring the {Sun}'s internal angular velocity from observed p-mode frequency splittings},
	journal = {ApJ}
}

@article{Clark1973,
	doi = {10.1017/s0022112073000340},
	url = {https://doi.org/10.1017/s0022112073000340},
	year = {1973},
	month = sep,
	publisher = {Cambridge University Press ({CUP})},
	volume = {60},
	number = {03},
	pages = {561},
	author = {Alfred Clark},
	title = {The linear spin-up of a strongly stratified fluid of small {Prandtl} number},
	journal = {JFM}
}

@article{Elliott1999,
	doi = {10.1086/307092},
	url = {https://doi.org/10.1086/307092},
	year = {1999},
	month = may,
	publisher = {American Astronomical Society},
	volume = {516},
	number = {1},
	pages = {475--481},
	author = {J. R. Elliott and D. O. Gough},
	title = {Calibration of the Thickness of the Solar Tachocline},
	journal = {ApJ}
}

@article{Featherstone2016a,
	doi = {10.3847/0004-637x/818/1/32},
	url = {https://doi.org/10.3847/0004-637x/818/1/32},
	year = {2016},
	month = feb,
	publisher = {American Astronomical Society},
	volume = {818},
	number = {1},
	pages = {32},
	author = {Nicholas A. Featherstone and Bradley W. Hindman},
	title = {The spectral amplitude of stellar convection and its scaling in the {high-Rayleigh-number} regime},
	journal = {ApJ}
}

@article{Ferraro1937,
	doi = {10.1093/mnras/97.6.458},
	url = {https://doi.org/10.1093/mnras/97.6.458},
	year = {1937},
	month = apr,
	publisher = {Oxford University Press ({OUP})},
	volume = {97},
	number = {6},
	pages = {458--472},
	author = {V. C. A. Ferraro},
	title = {The Non-uniform Rotation of the {Sun} and its Magnetic Field},
	journal = {MNRAS}
}

@article{ForgcsDajka2001,
	doi = {10.1023/a:1013389631585},
	url = {https://doi.org/10.1023/a:1013389631585},
	year = {2001},
	publisher = {Springer Science and Business Media {LLC}},
	volume = {203},
	number = {2},
	pages = {195--210},
	author = {E. Forg{\'{a}}cs-Dajka and K. Petrovay},
	title = {Tachocline Confinement by an Oscillatory Magnetic Field},
	journal = {SoPh}
}

@article{ForgcsDajka2002,
	doi = {10.1051/0004-6361:20020586},
	url = {https://doi.org/10.1051/0004-6361:20020586},
	year = {2002},
	month = jun,
	publisher = {{EDP} Sciences},
	volume = {389},
	number = {2},
	pages = {629--640},
	author = {E. Forg{\'{a}}cs-Dajka and K. Petrovay},
	title = {Dynamics of the fast solar tachocline. {I. Dipolar} field},
	journal = {A\&A}
}

@article{ForgcsDajka2004,
	doi = {10.1051/0004-6361:20031569},
	url = {https://doi.org/10.1051/0004-6361:20031569},
	year = {2004},
	month = jan,
	publisher = {{EDP} Sciences},
	volume = {413},
	number = {3},
	pages = {1143--1151},
	author = {E. Forg{\'{a}}cs-Dajka},
	title = {Dynamics of the fast solar tachocline. {II. Migrating field}},
	journal = {A\&A}
}

@article{Garaud2002,
	doi = {10.1046/j.1365-8711.2002.04961.x},
	url = {https://doi.org/10.1046/j.1365-8711.2002.04961.x},
	year = {2002},
	month = jan,
	publisher = {Oxford University Press ({OUP})},
	volume = {329},
	number = {1},
	pages = {1--17},
	author = {P. Garaud},
	title = {Dynamics of the solar tachocline. {I. An} incompressible study},
	journal = {MNRAS}
}

@article{Garaud2025,
  title = {Toward a Self-consistent Hydrodynamical Model of the Solar Tachocline},
  volume = {985},
  ISSN = {1538-4357},
  url = {http://dx.doi.org/10.3847/1538-4357/adc72b},
  DOI = {10.3847/1538-4357/adc72b},
  number = {2},
  journal = {ApJ},
  publisher = {American Astronomical Society},
  author = {Garaud,  P. and Gough,  D. O. and Matilsky,  L. I.},
  year = {2025},
  month = may,
  pages = {151}
}

@article{Gough1998,
	doi = {10.1038/29472},
	url = {https://doi.org/10.1038/29472},
	year = {1998},
	month = aug,
	publisher = {Springer Science and Business Media {LLC}},
	volume = {394},
	number = {6695},
	pages = {755--757},
	author = {D. O. Gough and M. E. McIntyre},
	title = {Inevitability of a magnetic field in the {Sun}'s radiative interior},
	journal = {Natur}
}

@article{Haynes1991,
	doi = {10.1175/1520-0469(1991)048<0651:otcoed>2.0.co;2},
	url = {https://doi.org/10.1175/1520-0469(1991)048<0651:otcoed>2.0.co;2},
	year = {1991},
	month = feb,
	publisher = {American Meteorological Society},
	volume = {48},
	number = {4},
	pages = {651--678},
	author = {P. H. Haynes and M. E. McIntyre and T. G. Shepherd and C. J. Marks and K. P. Shine},
	title = {On the `Downward Control' of Extratropical Diabatic Circulations by Eddy-Induced Mean Zonal Forces},
	journal = {JAtS}
}

@article{Howe2009,
	doi = {10.12942/lrsp-2009-1},
	url = {https://doi.org/10.12942/lrsp-2009-1},
	year = {2009},
	publisher = {Springer Science and Business Media {LLC}},
	volume = {6},
	pages = {1},
	author = {Rachel Howe},
	title = {Solar Interior Rotation and its Variation},
	journal = {LRSP}
}

@article{Kim2007,
	doi = {10.1051/0004-6361:20065971},
	url = {https://doi.org/10.1051/0004-6361:20065971},
	year = {2007},
	month = apr,
	publisher = {{EDP} Sciences},
	volume = {468},
	number = {3},
	pages = {1025--1031},
	author = {E.-J. Kim and N. Leprovost},
	title = {Self-consistent theory of turbulent transport 
	in the solar tachocline},
	journal = {A\&A}
}

@article{Korre2024,
	title = {On the Penetration of Large-scale Flows into Stellar Radiative Zones},
	volume = {964},
	ISSN = {1538-4357},
	url = {http://dx.doi.org/10.3847/1538-4357/ad2844},
	DOI = {10.3847/1538-4357/ad2844},
	number = {2},
	journal = {ApJ},
	publisher = {American Astronomical Society},
	author = {Korre,  Lydia and Featherstone,  Nicholas A.},
	year = {2024},
	month = mar,
	pages = {162}
}

@article{Kumar1997,
	doi = {10.1086/310477},
	url = {https://doi.org/10.1086/310477},
	year = {1997},
	month = feb,
	publisher = {American Astronomical Society},
	volume = {475},
	number = {2},
	pages = {L143--L146},
	author = {Pawan Kumar and Eliot J. Quataert},
	title = {Angular Momentum Transport by Gravity Waves and Its Effect on the Rotation of the Solar Interior},
	journal = {ApJ}
}

@article{Matilsky2022,
	doi = {10.3847/2041-8213/ac93ef},
	url = {https://doi.org/10.3847/2041-8213/ac93ef},
	year = {2022},
	month = nov,
	publisher = {American Astronomical Society},
	volume = {940},
	number = {2},
	pages = {L50},
	author = {Loren I. Matilsky and Bradley W. Hindman and Nicholas A. Featherstone and Catherine C. Blume and Juri Toomre},
	title = {Confinement of the Solar Tachocline by Dynamo Action in the Radiative Interior},
	journal = {ApJL}
}

@article{Matilsky2023,
	title = {The stellar thermal wind as a consequence of oblateness},
	volume = {526},
	ISSN = {1745-3933},
	url = {http://dx.doi.org/10.1093/mnrasl/slad121},
	DOI = {10.1093/mnrasl/slad121},
	number = {1},
	journal = {MNRASL},
	publisher = {Oxford University Press (OUP)},
	author = {Matilsky,  Loren I},
	year = {2023},
	month = aug,
	pages = {L100–L104}
}

@article{Matilsky2024,
	title = {Solar Tachocline Confinement by the Nonaxisymmetric Modes of a Dynamo Magnetic Field},
	volume = {962},
	ISSN = {1538-4357},
	url = {http://dx.doi.org/10.3847/1538-4357/ad18b2},
	DOI = {10.3847/1538-4357/ad18b2},
	number = {2},
	journal = {ApJ},
	publisher = {American Astronomical Society},
	author = {Matilsky,  Loren I. and Brummell,  Nicholas H. and Hindman,  Bradley W. and Toomre,  Juri},
	year = {2024},
	month = feb,
	pages = {189}
}

@article{Matilsky2025a,
  title = {Dynamo Confinement of a Radiatively Spreading Solar Tachocline Revealed by Self-consistent Global Simulations},
  volume = {991},
  ISSN = {2041-8213},
  url = {http://dx.doi.org/10.3847/2041-8213/adefe3},
  DOI = {10.3847/2041-8213/adefe3},
  number = {1},
  journal = {ApJL},
  publisher = {American Astronomical Society},
  author = {Matilsky,  Loren I. and Korre,  Lydia and Brummell,  Nicholas H.},
  year = {2025},
  month = sep,
  pages = {L1}
}

@article{Matilsky2026a,
  doi = {10.48550/ARXIV.2601.11943},
  url = {https://arxiv.org/abs/2601.11943},
  author = {Matilsky,  Loren I. and Korre,  Lydia and Brummell,  Nicholas H.},
  title = {A Dynamo Confinement Scenario for the Solar Tachocline and its Implications for Spin-down in the Radiative Spreading Regime},
  publisher = {arXiv},
  journal = {ApJ},
  year = {2026},
  copyright = {Creative Commons Attribution 4.0 International}
}

@article{Matsui2016,
	doi = {10.1002/2015gc006159},
	url = {https://doi.org/10.1002/2015gc006159},
	year = {2016},
	month = may,
	publisher = {American Geophysical Union ({AGU})},
	volume = {17},
	number = {5},
	pages = {1586--1607},
	author = {Hiroaki Matsui and Eric Heien and Julien Aubert and Jonathan M. Aurnou and Margaret Avery and Ben Brown and Bruce A. Buffett and Friedrich Busse and Ulrich R. Christensen and Christopher J. Davies and Nicholas Featherstone and Thomas Gastine and Gary A. Glatzmaier and David Gubbins and Jean-Luc Guermond and Yoshi-Yuki Hayashi and Rainer Hollerbach and Lorraine J. Hwang and Andrew Jackson and Chris A. Jones and Weiyuan Jiang and Louise H. Kellogg and Weijia Kuang and Maylis Landeau and Philippe Marti and Peter Olson and Adolfo Ribeiro and Youhei Sasaki and Nathanaël Schaeffer and Radostin D. Simitev and Andrey Sheyko and Luis Silva and Sabine Stanley and Futoshi Takahashi and Shin-ichi Takehiro and Johannes Wicht and Ashley P. Willis},
	title = {Performance benchmarks for a next generation numerical dynamo model},
	journal = {GGG}
}

@article{Mestel1987,
	doi = {10.1093/mnras/226.1.123},
	url = {https://doi.org/10.1093/mnras/226.1.123},
	year = {1987},
	month = may,
	publisher = {Oxford University Press ({OUP})},
	volume = {226},
	number = {1},
	pages = {123--135},
	author = {L. Mestel and N. O. Weiss},
	title = {Magnetic fields and non-uniform rotation in stellar radiative zones},
	journal = {MNRAS}
}

@article{Parker1955,
	doi = {10.1086/146087},
	url = {https://doi.org/10.1086/146087},
	year = {1955},
	month = sep,
	publisher = {American Astronomical Society},
	volume = {122},
	pages = {293},
	author = {Eugene N. Parker},
	title = {Hydromagnetic Dynamo Models.},
	journal = {ApJ}
}

@article{Spiegel1992,
	year = {1992},
	volume = {265},
	pages = {106},
	author = {E. A. Spiegel and J.-P. Zahn},
	title = {The solar tachocline},
	journal = {A\&A}
}

@article{Strugarek2011a,
	doi = {10.1002/asna.201111613},
	url = {https://doi.org/10.1002/asna.201111613},
	year = {2011},
	month = dec,
	publisher = {Wiley},
	volume = {332},
	number = {9-10},
	pages = {891--896},
	author = {A. Strugarek and A. S. Brun and J.-P. Zahn},
	title = {Magnetic confinement of the solar tachocline: The oblique dipole},
	journal = {AN}
}

@article{Wood2012,
	doi = {10.1088/0004-637x/755/2/99},
	url = {https://doi.org/10.1088/0004-637x/755/2/99},
	year = {2012},
	month = aug,
	publisher = {American Astronomical Society},
	volume = {755},
	number = {2},
	pages = {99},
	author = {T. S. Wood and N. H. Brummell},
	title = {Transport by meridional circulations in solar-type stars},
	journal = {ApJ}
}

@article{Wood2018,
	doi = {10.3847/1538-4357/aaa6d5},
	url = {https://doi.org/10.3847/1538-4357/aaa6d5},
	year = {2018},
	month = jan,
	publisher = {American Astronomical Society},
	volume = {853},
	number = {2},
	pages = {97},
	author = {T. S. Wood and N. H. Brummell},
	title = {A Self-consistent Model of the Solar Tachocline},
	journal = {ApJ}
}

@Software{Featherstone2021,
	author = "{Featherstone}, N.~A. and {Edelmann}, P.~V.~F. and {Gassmoeller}, R. and {Matilsky}, L.~I. and {Orvedahl}, R.~J. and {Wilson}, C.~R.",
	title="Rayleigh 1.0.1",
	year="2021",
	organization="",
	optkeywords="Rayleigh",
	doi="10.5281/zenodo.1158289",
	opturl="https://doi.org/10.5281/zenodo.5774039"
}

@INPROCEEDINGS{Matilsky2021,
	doi = {10.5281/ZENODO.4750777},
	url = {https://zenodo.org/record/4750777},
	author = {Matilsky,  Loren Isaac and Toomre,  Juri},
	title = {Building and maintaining a solar tachocline through convective dynamo action},
	booktitle = {20.5th Cambridge Workshop on Cool Stars, Stellar Systems, and the Sun},
	publisher = {Zenodo},
	year = {2021},
	editor = {{Wolk}, S.~J.},
	copyright = {Creative Commons Attribution 4.0 International}
}

\end{document}